\documentclass[11pt,a4paper]{article}
\pdfoutput=1
\usepackage{jcappub}
\usepackage{verbatim}
\usepackage{placeins}

\title{Constraining ultralight scalar dark matter in the Galactic Center with the S2 orbit}

\author[a,b]{Jiang-Chuan Yu,}
\author[c]{Yan Cao}
\author[b,d,1]{and Lijing Shao}
\note[1]{Corresponding author.}

\affiliation[a]{Department of Astronomy, School of Physics, Peking University,
Beijing 100871, China}
\affiliation[b]{Kavli Institute for Astronomy and Astrophysics, Peking University,
Beijing 100871, China}
\affiliation[c]{School of Physics, Nanjing University, Nanjing 210093, China}
\affiliation[d]{National Astronomical Observatories, Chinese Academy of Sciences,
Beijing 100101, China}

\emailAdd{jcyu25@stu.pku.edu.cn}
\emailAdd{yancao@smail.nju.edu.cn}
\emailAdd{lshao@pku.edu.cn}

\abstract{
The dense environment of our Galactic Center (GC) offers a unique laboratory for probing ultralight dark matter (ULDM). We explore the prospect of detecting a scalar ULDM field through its effects on the orbital dynamics of S-stars around the supermassive black hole in the GC, Sgr~A$^*$. We consider both linear and quadratic couplings between the real scalar field $\phi$ and Standard Model particles, and analyze two representative ULDM structures: the scalar gravitational atom and the spherical soliton. We first find that quadratic coupling induces a non-oscillatory perturbation, leading to a long-term secular orbital evolution. We use the observed periastron precession rate of S2 star to put stringent constraints on the total ULDM mass in the GC and the quadratic coupling constant. For the gravitational atom $|211\rangle$ state, assuming an equatorial S2 orbit, we constrain the mass ratio of ULDM to Sgr~A$^*$ to $\beta
\lesssim 10^{-3}$ at $m \sim 10^{-18}$ eV, and for the spherical soliton which extends to $\sim 0.2\,$pc, the mass ratio is limited to $\beta \lesssim 1$ at $m \sim 3\times10^{-20}$ eV. Notably, over part of the mass range \(10^{-20}\,\mathrm{eV}\lesssim m\lesssim10^{-18}\,\mathrm{eV}\), the S2 constraints are stronger than the corresponding bounds from Cassini and MICROSCOPE.
}

\arxivnumber{2604.08053}

\begin{document}
\maketitle
\flushbottom

\section{Introduction}
Ultralight scalar bosons have emerged as one of the most compelling dark matter (DM) candidates, arising naturally in many extensions of the Standard Model (SM)~\cite{Peccei:1977hh,Weinberg:1977ma,Wilczek:1977pj,Dine:1982ah,Hui:2016ltb,Marsh:2015xka,Agrawal:2021dbo,eolchinski:1998rr,Kim:1986ax,Fayet:1990wx,Svrcek:2006yi,Arvanitaki_2010,Arias:2012az,Lyth:1998xn,Kolb:2020fwh,Khlopov:1985fch}. From a low-energy effective-field-theory (EFT) perspective, the leading interactions of a light scalar field $\phi$ with the SM fields can take the form of linear and quadratic couplings. Linear couplings, represented schematically as $\phi\,\mathcal{O}_{\mathrm{SM}}$, are expected to be the most straightforward interactions, with representative realizations such as the dilaton model~\cite{Damour:2010rp,Damour:2010rm}. Instead of $\phi\,\mathcal{O}_{\mathrm{SM}}$, quadratic couplings take the form $\phi^{2}\,\mathcal{O}_{\mathrm{SM}}$, and the scalar field respects the $\phi \to -\phi$ symmetry. Relevant models include, for example, QCD axion interactions~\cite{Okawa:2021fto,Bauer:2023czj,cheng2025backgroundenhancedaxionforceaxion}.

While the nature of DM is not understood yet, ultralight DM (ULDM) is a prominent model. Due to its oscillatory nature, ULDM induces periodic variations in the SM coupling constants and particle masses. These variations can be detected by high-precision instruments such as atomic clocks~\cite{Derevianko:2013oaa,Arvanitaki:2014faa,VanTilburg:2015oza,Hees:2016gop}, laser interferometers~\cite{Stadnik:2014tta,Yu:2023iog}, resonant-mass detectors~\cite{Arvanitaki:2015iga}, and radio telescopes~\cite{Caputo:2019tms, An:2022hhb}. Additionally, scalar interactions will lead to a ``fifth force'', which can cause apparent deviations from the General Relativity (GR) in the gravity sector. If these interactions are non-universal, the resulting composition-dependent forces could violate the Weak Equivalence Principle (WEP) and break the Universality of Free Fall (UFF). These unique theoretical predictions open an avenue to a variety of complementary experimental approaches. In a universal interaction scenario, detection methods include binary orbital resonance~\cite{Blas:2016ddr,Blas:2019hxz}, pulsar timing arrays (PTA)~\cite{Porayko:2014rfa, Smarra:2024kvv}, space-based gravitational-wave detectors~\cite{Gasparotto:2025pms}, and the Cassini mission~\cite{Bertotti:2003rm,Armstrong:2003ay}. In the case of composition-dependent interactions, UFF experiments, such as torsion-balance measurements~\cite{Smith:1999cr,Schlamminger:2007ht,Wagner:2012ui} and the MICROSCOPE mission~\cite{MICROSCOPE:2022doy}, were used to probe the existence of fifth force. Recent LAGEOS II pericentre-precession measurements provide complementary constraints on quadratically coupled ULDM, particularly at strong coupling~\cite{Burrage:2026lageos}.

The Galactic Center (GC) provides a unique natural ULDM laboratory, since dense bound structures such as superradiant cloud~\cite{PhysRevD.22.2323,Cardoso:2005vk,PhysRevD.76.084001,Arvanitaki_2010,PhysRevD.87.043513,Brito:2014wla,Brito:2015oca,B1,ZJ,B2,B3} or solitonic core~\cite{Schive:2014dra,Schive:2014hza,Chavanis:2019bnu,Bar:2019pnz,Davies_2020,Annulli:2020lyc,Zagorac:2022xic,Aghaie:2023lan,liao2025decipheringsolitonhalorelationfuzzy,Sarkar:2025tiy} can form there. If present, these configurations will induce dynamical perturbations on nearby stars, providing an exceptional opportunity to probe ULDM. Over the past two decades, high-precision astrometric and spectroscopic monitoring of the S-cluster stars orbiting Sgr A* has enabled direct and stringent tests of GR and fundamental physics~\cite{GRAVITY:2020gka,GRAVITY:2021xju,GRAVITY:2024tth}. The recent discovery of S301, whose orbit is expected to be directly sensitive to the spin of Sgr~A$^*$, further highlights the potential of S-stars as precision probes of the GC~\cite{Dayem:2026ktt}. In recent years, numerous studies have used orbital data of the S-cluster stars to investigate the distribution of DM at the GC~\cite{Lacroix:2018zmg,Shen:2023kkm,Zakharov:2007fj,Heissel:2021pcw,GRAVITY:2019tuf,Yuan:2022nmu,DellaMonica:2022kow,GRAVITY:2023cjt,GRAVITY:2023azi,bai2025probingaxionsspectroscopicmeasurements,tomaselli2025probingdenseenvironmentssgr,Bar:2019pnz,Chan:2022nra} and to probe possible DM-SM non-gravitational interactions~\cite{acevedo2025darkdragsagittariusa,gustafson2025probingdarkmatterinteractions}. Here our work gives a first exploration of the scenario where ordinary matter is directly coupled to the background ULDM using S stars.

In this study, we use precision observations of the S2 stellar orbit to detect the ULDM-SM interactions and to set constraints on the coupling constant as well as the total mass of the ULDM structures in the GC. The paper is organized as follows. In Sec.~\ref{sec:model}, we introduce two types of non-gravitational interactions, namely universal and non-universal couplings, between the scalar field and the baryonic matter composing the stars. Then in Sec.~\ref{sec:acc}, we derive the orbital dynamics of S-stars, affected by the extra acceleration due to gravitational and non-gravitational couplings in both scalar cloud and soliton scenarios. Constraints derived from observations of the orbital precession of S2 are presented in Sec.~\ref{sec:con}. Finally, we summarize our results in Sec.~\ref{sec:sum}. Throughout this paper, we adopt the flat spacetime metric $\eta_{ab} = \mathrm{diag}(1,-1,-1,-1)$ and natural units where $\hbar = c = G = 1$.

\section{ULDM-Matter Coupling}\label{sec:model}

We consider the ULDM described by a canonical \textit{real} scalar field $\phi$ that is minimally coupled to gravity, with scalar mass $m$. Throughout the main text, we set scalar self-interactions to zero; Appendix~\ref{app:self-interaction} assesses the validity of this approximation for a quartic self-interaction. The scalar boson may additionally interact with the SM particles. Here we are interested in the possible modification of the \textit{inertial mass} of a macroscopic baryonic object in a \textit{classical} background of the scalar field,
\begin{align}
   \mathcal{M}(\phi) = \mathcal{M}_{0}[1+\Theta(\phi)], \label{modified_mass}
\end{align}
with $\mathcal{M}_{0}$ being the mass in the absence of this background. For example, in the case of a ``universal coupling'', the function $\Theta(\phi)$ takes the universal form,
\begin{equation}
\Theta(\phi)= \frac{\phi}{\Lambda_1} + \frac{\phi^2}{2\Lambda_2^2} + \mathcal{O}(\phi^3),\label{Theta}
\end{equation}
for all objects, where $\Lambda_1$, $\Lambda_2$ are the energy scales of new physics generating the linear and quadratic couplings, respectively. The mass modifications may also be non-universal. As a concrete example, we consider the parameterized dilaton model, with the interaction Lagrangian given by \cite{Damour:2010rp,Damour:2010rm},
{\small
\begin{align}
\mathcal{L}_{\text{int}}={}&\sqrt{4\pi}\,\phi
\left[-\frac{d_{g}^{(1)}\beta_{3}}{2g_{3}}
G^{A}_{\mu \nu}G^{A \mu \nu}
-\sum_{i=u,d}
(d_{m_{i}}^{(1)}+\gamma_{m_{i}}d_{g}^{(1)})m_{i}\bar{\psi}_{i}\psi_{i}\right]\nonumber\\
&+\frac{4\pi\phi^2}{2}
\left[-\frac{d_{g}^{(2)}\beta_{3}}{2g_{3}}
G^{A}_{\mu \nu}G^{A \mu \nu}
-\sum_{i=u,d}
(d_{m_{i}}^{(2)}+\gamma_{m_{i}}d_{g}^{(2)})m_{i}\bar{\psi}_{i}\psi_{i}\right],
\end{align}
}
where $G^{A}_{\mu \nu}$ is the gluon field strength tensor, $g_{3}$ is the $\mathrm{SU(3)}$ gauge coupling, $\beta_{3}$ is the QCD beta function, $\gamma_{m_{u,d}}$ are the anomalous dimensions of the $u$ and $d$ quarks, and $\{d_{g}, d_{m_{u}}, d_{m_{d}} \}$ are the {coupling parameters} to the gluon field and the quark masses, respectively. The index ``1'' refers to linear coupling and ``2'' refers to quadratic coupling. The resulting mass modulation is~\cite{Damour:2010rp,Damour:2010rm}
\begin{align}
\Theta(\phi)\simeq \sqrt{4\pi}\,d_g^{*(1)}\,\phi + \frac{4\pi d_g^{*(2)}}{2}\,\phi^2,
\end{align}
where
\begin{align}
    d_{g}^{*(i)} &\simeq d_{g}^{(i)} + 0.093 \big(d_{\hat{m}}^{(i)}-d_{g}^{(i)} \big), \\
    d_{\hat{m}}^{(i)} &\equiv \frac{d_{m_{d}}^{(i)}m_{d}+d_{m_{u}}^{(i)}m_{u}}{m_{d}+m_{u}}.
\end{align}

\section{Stellar orbital dynamics in a scalar cloud}\label{sec:acc}

The scalar boson may form a dense classical condensate around an astrophysical black hole. We consider a small point-like test body traveling in the scalar cloud, whose mass is replaced by the ``dynamical mass'', $\mathcal{M}=\mathcal{M}_0[1+\Theta(\phi)]$, according to Eq.~\eqref{modified_mass}. Its worldline action reads
\begin{equation}
S=-\int d\tau\,\mathcal{M}\sqrt{g_{ab}\dot x^a\dot x^b},
\end{equation}
where $\dot x^a=dx^a/d\tau$ is the four-velocity, with $\tau$ being the proper time. The resultant four-acceleration of the body is
\begin{equation}
a^a\equiv\ddot{x}^a+\Gamma_{bc}^a\dot{x}^b\dot{x}^c=(g^{ab}-\dot x^a\dot x^b)\frac{\partial_b\mathcal{M}}{\mathcal{M}}.
\end{equation}
In the flat-spacetime, slow-motion limit, $\dot x^a\approx (1,\mathbf{v})$, thus
\begin{equation}
a^i=(g^{ib}-\dot x^i\dot x^b)\frac{\partial_b\mathcal{M}}{\mathcal{M}}\approx -\frac{\partial_t\mathcal{M}}{\mathcal{M}}v^i-\frac{\partial_i\mathcal{M}}{\mathcal{M}}.
\end{equation}
At the Newtonian level, the metric is approximated by $g_{ab}dx^a dx^b=(1+2\Phi)dt^2-(1-2\Phi)|d\mathbf{x}|^2$, with the Newtonian potential $|\Phi|\ll 1$, hence $\Gamma^i_{bc}\dot x^b\dot x^c\approx \Gamma^i_{00}\approx \partial_i\Phi$. 

In the nonrelativistic regime, we treat the scalar cloud as a solution to the Schr\"odinger--Poisson (SP) equation with $\Phi=\Phi_\text{BH}+\Phi_\text{c}$, where $\Phi_\text{BH}=-M/r$, $M$ is the BH mass, and $\Phi_\text{c}$ is the Newtonian potential sourced by the cloud (see Appendix~\ref{app:self-interaction}). In the absence of the cloud, the geodesic equation can be organized into a post-Newtonian (PN) expansion in the harmonic coordinates~\cite{Will:2016pgm}. Since we are interested in the regime where both the relativistic corrections and the effects of the cloud are small, the perturbing acceleration of the test body can be approximated by a sum of the PN acceleration, $\delta \mathbf{a}_\text{PN}^\text{(vac)}$, in vacuum and the Newtonian acceleration due to the unperturbed cloud,
\begin{align}
\mathbf{a} &\equiv \frac{d\mathbf{v}}{dt}=-\frac{M}{r^2}\mathbf{e}_r+\delta\mathbf{a}, \label{orbital_dynamics}
\\
\delta \mathbf{a}&=\delta \mathbf{a}_\text{g} + \delta \mathbf{a}_\text{n} + \delta \mathbf{a}_\text{PN}^\text{(vac)},\\[3pt]
\delta \mathbf{a}_\text{g}&=-\nabla\Phi_\text{c},\\
\delta \mathbf{a}_\text{n}&=-\frac{\partial_t\mathcal{M}}{\mathcal{M}} \mathbf{v}-\frac{\nabla \mathcal{M}}{\mathcal{M}}\approx-\frac{d\Theta}{d\phi}(\mathbf{v}\,\partial_t\phi+\nabla \phi), \label{noneq}
\end{align}
assuming that $|\Theta|\ll1$. We neglect the gravitational backreaction of the cloud perturbation induced by the small body (which is expected to be much weaker than the conservative effects if the cloud is only weakly perturbed~\cite{Cao:2024wby}), as we focus on orbital evolution on the orbital-period time scale.

In a nonrelativistic bound state, the real scalar field $\phi$ locally oscillates at a frequency $\omega_\phi \sim m$; for $m> 10^{-21}\,\text{eV}$, this is much higher than the orbital frequency $2\pi/T\sim 10^{-23}\,\text{eV}$ of the S2 star. In the case of a linear coupling\footnote{For the linear coupling, the small object effectively carries a scalar charge. Given that the central BH is assumed to possess no scalar charge, 
the magnitude of the scalar self-force $\sim (m_* M/r^2)\times (m_*/M)v_*^3/\Lambda_1^2$, where $v_*$ and $m_*$ are respectively the orbital velocity and mass of the object. It is negligible for the S2 star given the constraint $\Lambda_1^{-1}\lesssim 10^{-21}\,\text{GeV}^{-1}$~\cite{Blas:2016ddr}.} ($\Theta\propto \phi$), the resulting $\delta \mathbf{a}_\text{n}$ also undergoes such a rapid oscillation, and its secular effect is therefore suppressed. The metric perturbation sourced by the cloud should also {contain} oscillatory components with temporal frequencies $\sim \mathcal{O}(m)$, whose effect is likewise negligible for the same reason. In the case of a quadratic coupling, however, $\delta \mathbf{a}_\text{n}$ may contain a non-oscillatory component given by $\overline{\delta \mathbf{a}}_\text{n} \equiv (\omega_\phi/2\pi)\int_0^{2\pi/\omega_\phi} dt\, \delta \mathbf{a}_\text{n}$, where the integrand is evaluated at a fixed spatial position. The effect of this component persists during the orbital time scale. We parameterize the quadratic coupling by $\Theta(\phi)=\frac{1}{2}C\phi^2$, e.g., $C = (1/\Lambda_2)^2$ for a universal coupling, and $C \approx 4\pi d_g^{*(2)}$ in the dilaton model. The acceleration in Eq.~\eqref{noneq} treats S2 as a point particle. At sufficiently large positive quadratic coupling, however, the scalar field is suppressed inside the extended star, reducing its effective scalar charge and hence $\overline{\delta\mathbf{a}}_\mathrm{n}$. We include this finite-size response for all constraints below through $\overline{\delta\mathbf{a}}_\mathrm{n}^{\mathrm{FS}}=\mathcal{A}_\star\overline{\delta\mathbf{a}}_\mathrm{n}$, using a uniform-density model with $M_\star=13.6\,M_\odot$ and $R_\star=5.53\,R_\odot$. The derivation of $\mathcal{A}_\star$ is given in Appendix~\ref{app:finite-size}. In what follows, we consider two types of scalar clouds, the gravitational atom (GA) and the spherical soliton, and present the explicit forms of $\delta \mathbf{a}_\text{g}$ and $\overline{\delta \mathbf{a}}_\text{n}$.

\subsection{Gravitational atom}

A gravitational atom in the $|211\rangle$ state populated by the BH superradiance corresponds to a scalar field profile,
\begin{equation}\label{phibo}
\phi(t,\mathbf{x})\approx-\frac{\alpha ^2 \sqrt{\beta } }{4 \sqrt{2 \pi }}e^{-x/2} x \sin \theta\, \cos \left[m\left(1-\frac{\alpha^2}{8}\right) t- \varphi\right],
\end{equation}
in the spherical coordinates $\{r,\theta,\varphi\}$, with the $z$-axis aligned with the BH spin. Here $\alpha \equiv m M$ is the gravitational fine-structure constant, $\beta \equiv M_\text{c}/M$ is the cloud-to-BH mass ratio, $x \equiv r/r_\text{c}$, and $r_\text{c}\equiv M/\alpha^2$ is the gravitational Bohr radius. The nonrelativistic wavefunction associated with Eq.~\eqref{phibo} is not an exact solution of the SP equation, but provides a good approximation in the regime $\beta\ll 1$. For sufficiently small $\alpha$, the superradiant growth timescale may exceed the Galactic age~\cite{Brito:2015oca}. In this regime, we treat the $|211\rangle$ configuration as a phenomenological GA profile. The gravitational acceleration $\delta \mathbf{a}_{\text{g}}=-\nabla\Phi_\text{c}$ is given by (see for example \cite{yu2025detectingultralightdarkmatter})
\begin{align}
\frac{\delta \mathbf{a}_{\text{g}}}{\beta \alpha^4/M}={}&
\bigg\{-\frac{1}{x^2}+\frac{e^{-x}}{16x^4}\Big[
144-144e^x+144x+88x^2+40x^3+14x^4+4x^5+x^6\nonumber\\
&\quad+\left(-432+432e^x-432x-216x^2-72x^3-18x^4-4x^5-x^6\right)
\cos^2\theta\Big]\bigg\}\mathbf{e}_r\nonumber\\
&+\frac{e^{-x}}{8x^4}\Big[
\left(-144+144e^x-144x-72x^2-24x^3-6x^4-x^5\right)
\cos \theta \sin \theta\Big]\mathbf{e}_{\theta} .
\end{align}
Meanwhile, the quadratic coupling gives rise to a non-oscillatory acceleration,
\begin{align}
\frac{\overline{\delta \mathbf{a}}_\text{n}}{\beta \alpha^4/M}
={}&\frac{\alpha^2C}{64\pi}\left\{
e^{-x}\left(\frac{x}{2}-1\right)x\sin^2\theta\,\mathbf{e}_r
-e^{-x}x\sin\theta \cos\theta\,\mathbf{e}_{\theta}\right\},
\end{align}
due to the spatial gradient of the scalar field.

\subsection{Spherical soliton}

The spherical soliton, i.e., the spherically symmetric ground state, corresponds to the scalar field profile (see Ref.~\cite{yu2025detectingultralightdarkmatter} and Appendix~\ref{app:self-interaction}),
\begin{equation}
\phi(t,\mathbf{x})\approx \sqrt{\frac{2}{m}}\,h(r)\,\cos \big[(m+E)t\big].
\end{equation}
The radial profile $h(r)$, energy level $E$ and the gravitational acceleration can be computed numerically, as presented in Ref.~\cite{yu2025detectingultralightdarkmatter}. The non-oscillatory acceleration due to the quadratic coupling is given by
\begin{align}
 \overline{\delta \mathbf{a}}_\text{n} &= -\frac{Ch}{m}\left(\frac{d h}{d r}\right)\mathbf{e}_r.
\end{align}

\section{Constraints from S2 orbit}\label{sec:con}

The acceleration induced by the scalar ULDM contributes, in addition to GR effects, to the periastron precession of S-star orbits. The perturbing acceleration in Eq.~\eqref{orbital_dynamics} can be expressed by
\begin{equation}
    \delta{\mathbf{a}}  = \mathcal{R}\,\mathbf{e}_r + \mathcal{S}\,\mathbf{e}_\lambda + \mathcal{W}\, \textbf{e}_Z,
\end{equation}
where $\textbf{e}_Z$ points along $\mathbf{r}\times\mathbf{v}$, and $\mathbf{e}_\lambda=\mathbf{e}_Z \times \mathbf{e}_r$. The evolution of the longitude of periastron $\omega$ of the instantaneous osculating Keplerian orbit with the total mass parameter $M_\text{tot}\approx M$ is given by~\cite{Poisson_Will_2014}
\begin{align}\label{m5}
\frac{d\omega}{dt}={}&\frac{1}{e}\sqrt{\frac{a(1-e^2)}{M}}
\left[-\cos f\,\mathcal{R}
+\frac{2+e\cos f}{1+e\cos f}\sin f\,\mathcal{S}
-e\cot i\,\frac{\sin(\omega+f)}{1+e\cos f}\mathcal{W}\right],
\end{align}
where $f$ is the true anomaly in the osculating orbit. If $\delta{\mathbf{a}}$ is sufficiently small, the accumulated change of $\omega$ over one orbital period {$T=2\pi\sqrt{a^3/M}$} is approximated by $\Delta \omega\approx\int_0^T dt\,\frac{d\omega}{dt}$, with the osculating elements in the integrand fixed to constant values.

The S-cluster comprises a population of stars on tight orbits around Sgr A*. Within the S-cluster, S2 is the star whose Schwarzschild precession is constrained with the highest observational accuracy. The orbital parameters of S2, including its semi-major axis \( a = 5.016  \, \text{mpc} \) and eccentricity \( e = 0.884 \), are well determined in Ref.~\cite{GRAVITY:2024tth}. For simplicity, we assume an orbital plane orthogonal to the black hole spin.

 We use the latest measurements of S2 provided by the GRAVITY collaboration~\cite{GRAVITY:2024tth}, which gives a measurement on the periastron advance {$\Delta \omega \in12.1'\times [0.918-0.128,\,0.918+0.128]$}. {According to Eq.~(\ref*{orbital_dynamics}), the periastron advance can be approximated by $\Delta\omega\approx\Delta\omega_\mathrm{S}+\Delta\omega_\mathrm{g}+\Delta\omega_\mathrm{n}$, where $\Delta\omega_\mathrm{S}=6\pi M/[a(1-e^2)]$ is the Schwarzschild contribution, and $\Delta\omega_\mathrm{g}$ and $\Delta\omega_\mathrm{n}$ denote the contributions accumulated over one orbit from $\delta\mathbf{a}_\mathrm{g}$ and $\overline{\delta\mathbf{a}}_\mathrm{n}$, respectively.} In deriving these constraints, we neglect additional contributions to the gravitational potential in the Galactic Center, such as that from an extended mass distribution around Sgr~A$^*$~\cite{Hees:2017aal,GRAVITY:2024tth}.
For the non-gravitational effect of ULDM, only the non-oscillatory effect contributes significantly to the long-term precession. Therefore, we focus on setting constraints for the case of quadratic coupling, with the relevant parameters being \( \big\{\alpha, \beta, \Lambda_2\,\text{or}\,d_g^{(2)} \big\} \), where \( \alpha \) and \( \beta \) 
relate to the particle mass of the ULDM and the total ULDM mass in the GC, and $\Lambda_2\,\text{or}\,d_g^{(2)}$ is the coupling constant.

\begin{figure}[t]
\centering
\includegraphics[width=0.62\textwidth]{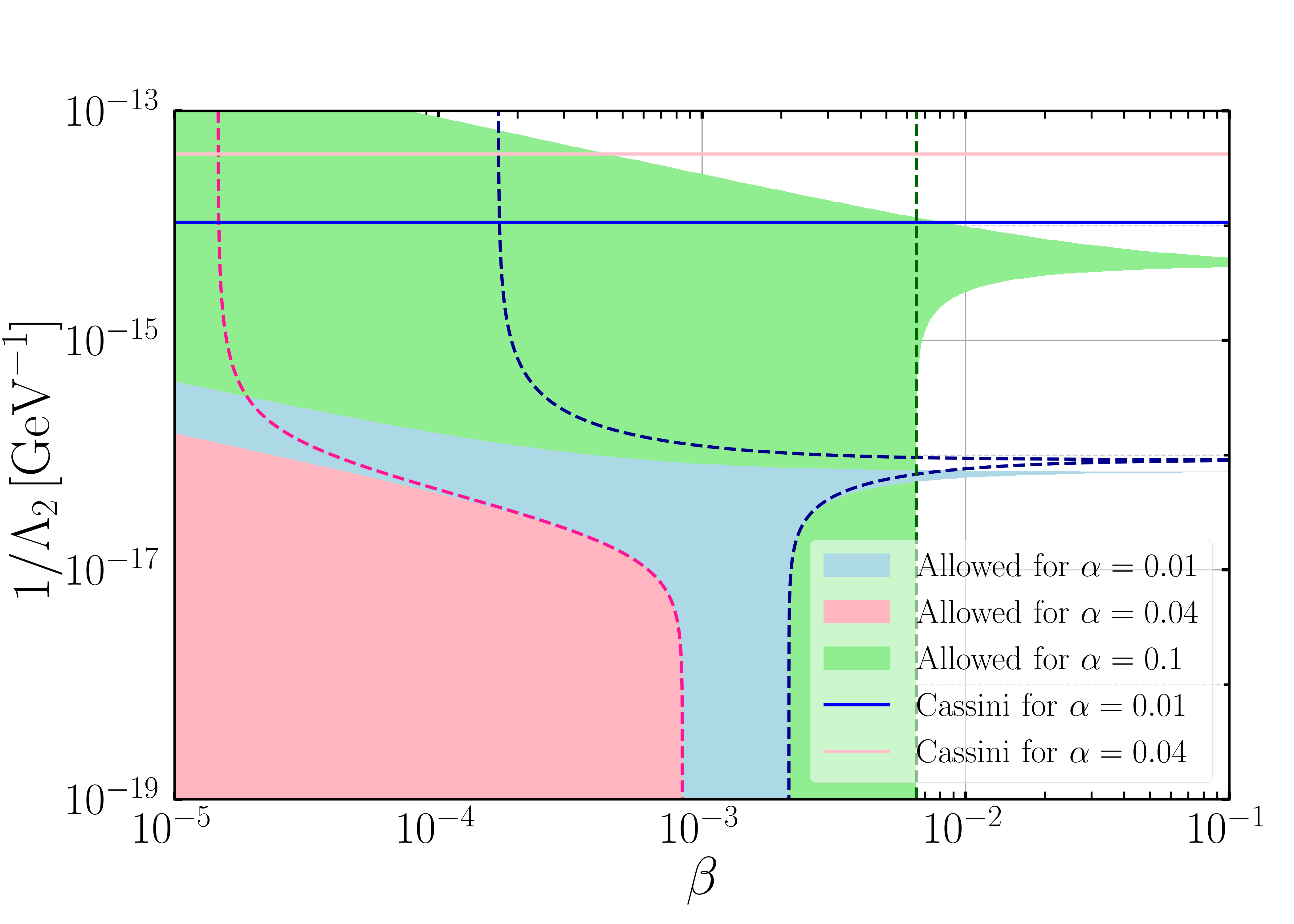}\par
\vspace{1mm}
\includegraphics[width=0.62\textwidth]{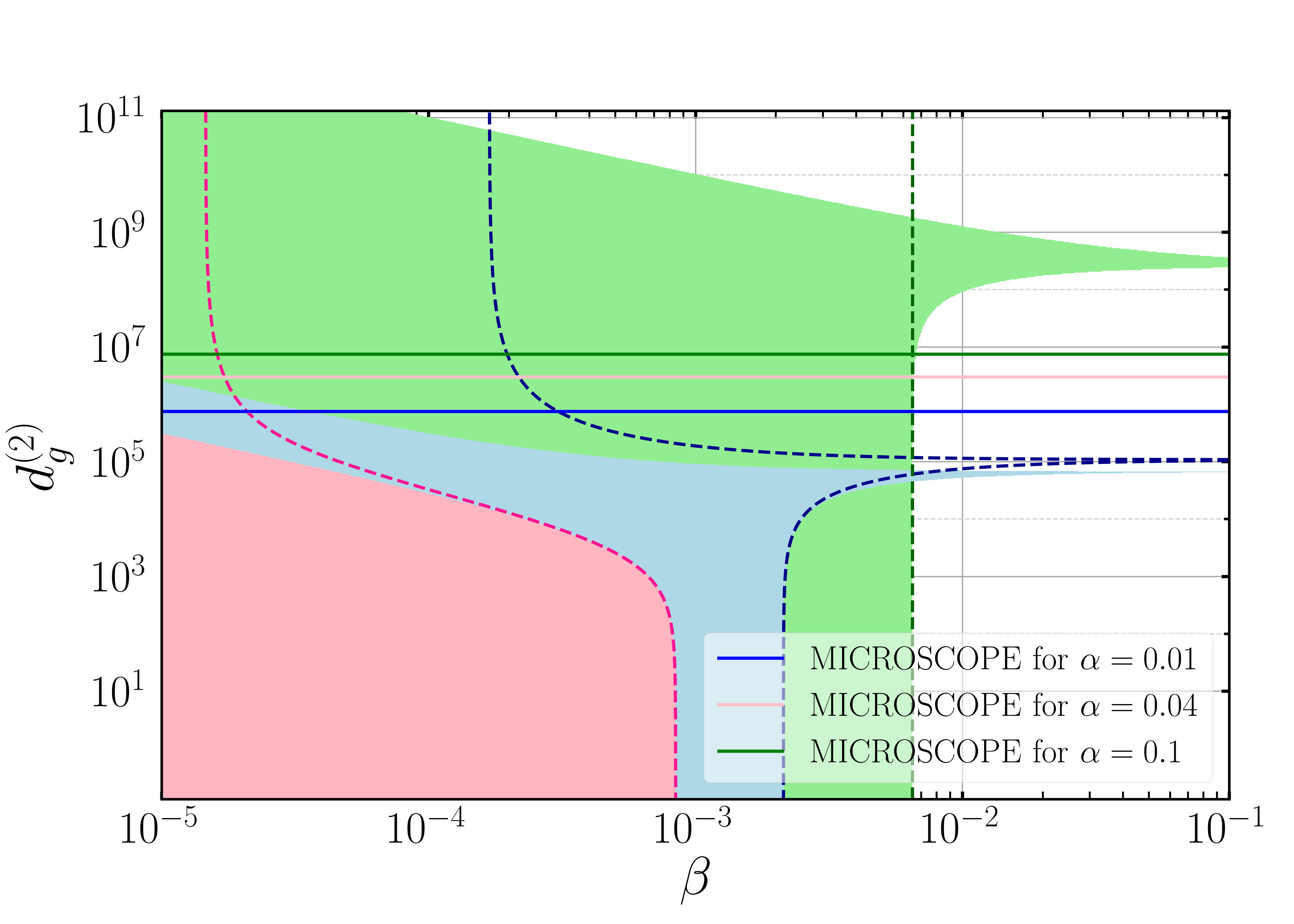}
\caption{Constraints on $1/\Lambda_2$ (top) and $d_g^{(2)}$ (bottom) versus $\beta$ in the cases of $\alpha=0.01,\,0.04,\,0.1$ for the $|211\rangle$ state GA, presented by blue, pink, and green regions, respectively. {The shaded regions show the parameter space allowed at the 1-$\sigma$ observational level in the point-particle treatment, while the darker blue, pink, and green dashed contours mark the corresponding allowed-region boundaries after including the finite-size response of S2.} Constraints from Cassini stochastic gravitational waves measurement \cite{Armstrong:2003ay} and MICROSCOPE experiment \cite{MICROSCOPE:2022doy} are also {shown} for comparison. {The regions above the solid comparison lines are excluded.}}
\label{fig:ga}
\end{figure} 
 
 The 1-$\sigma$ constraints for the \( |211\rangle \) state GA are shown in Fig.~\ref{fig:ga}. The top panel displays the 2D constraints on the universal coupling constant \( 1/\Lambda_2 \) and the mass ratio \( \beta \), with \( \alpha \) fixed at values of \( 0.01 \), \( 0.04 \), and \( 0.1 \). The bottom panel shows the case for the dilaton model. For comparison, we also include the constraints from (top panel) the Cassini stochastic gravitational wave background constraint \cite{Armstrong:2003ay} and (bottom panel) the MICROSCOPE experiment \cite{MICROSCOPE:2022doy}. To directly compare with the MICROSCOPE results, we set all coupling parameters, except for \( d_g^{(2)} \), to zero. {The shaded regions show the allowed parameter space in the point particle treatment, while the dashed contours include the finite size response.}

{In the point particle approximation,} we find that for $\alpha \sim 0.04$ ($m \sim 10^{-18}$ eV), the parameter region with $\beta \gtrsim 10^{-3}$ and $1/\Lambda_2 \gtrsim 10^{-16} \, \text{GeV}^{-1}$ is excluded, yielding constraints that are at least two orders of magnitude tighter than those obtained from Cassini. This is attributed to the combined effects of gravitational and non-gravitational interactions where gravitational interactions dominate for larger $\beta$ while non-gravitational interactions are more significant for smaller $\beta$, leading to joint constraints on both parameters. {With the finite size response, the S2 limit remains stronger than Cassini for $\beta\gtrsim1.5\times10^{-5}$.} {For the same value $\alpha=0.04$, the finite size corrected limit on $d_g^{(2)}$ remains stronger than MICROSCOPE for $\beta\gtrsim2\times10^{-5}$.} For $\alpha \sim 0.01$ and $\alpha \sim 0.1$, a narrow allowed parameter region persists at larger $\beta$ due to the opposite signs of the contributions to the periastron precession induced by $\delta \mathbf{a}_\text{g}$ and $\overline{\delta \mathbf{a}}_\text{n}$. {For the adopted S2 model, $\mathcal{A}_\star$ is reduced by $10\%$ at $1/\Lambda_2\simeq4\times10^{-17}\,\mathrm{GeV}^{-1}$ or $d_g^{(2)}\simeq2\times10^4$. The combination $C\mathcal{A}_\star$ approaches a constant and reaches $90\%$ of its asymptotic value at $1/\Lambda_2\simeq7\times10^{-16}\,\mathrm{GeV}^{-1}$ or $d_g^{(2)}\simeq7\times10^6$. At strong coupling, the induced non gravitational precession therefore becomes nearly coupling independent, producing the vertical limiting behavior of the dashed boundaries.}

As $\alpha$ increases, the Bohr radius $r_\text{c} = M/\alpha^2$ of the GA decreases, increasing the enclosed mass within the orbit and thereby tightening the constraints. However, for a sufficiently large $\alpha$, the constraints weaken as the density around the orbit drops. {Consequently, the overlap between the increasingly compact GA and the S2 orbit first increases and then decreases.} This leads to the strongest sensitivity near $\alpha\simeq0.04$, where the characteristic cloud scale is comparable to the orbital scale of S2.

\begin{figure}[t]
\centering
\includegraphics[width=0.62\textwidth]{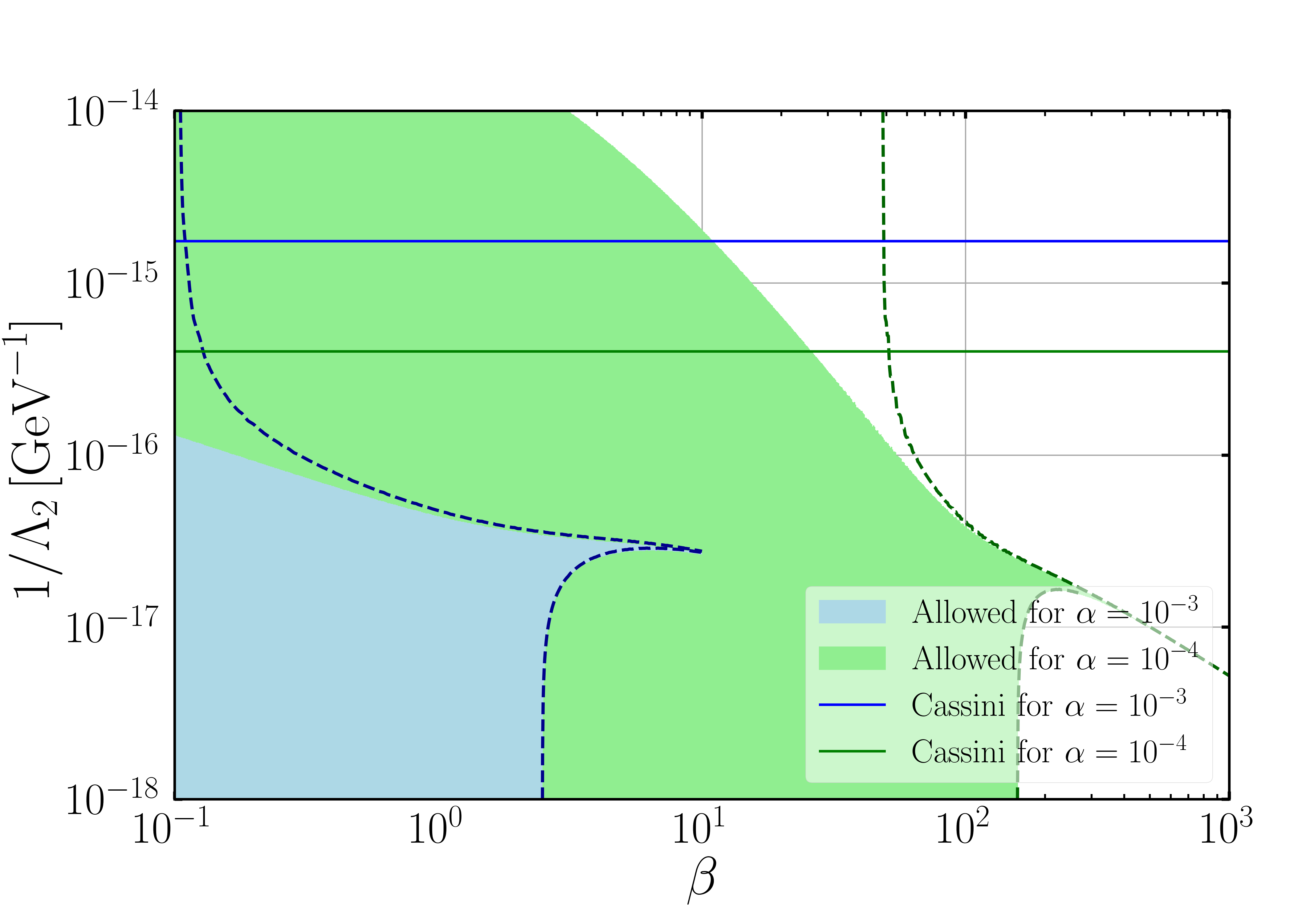}\par
\vspace{1mm}
\includegraphics[width=0.62\textwidth]{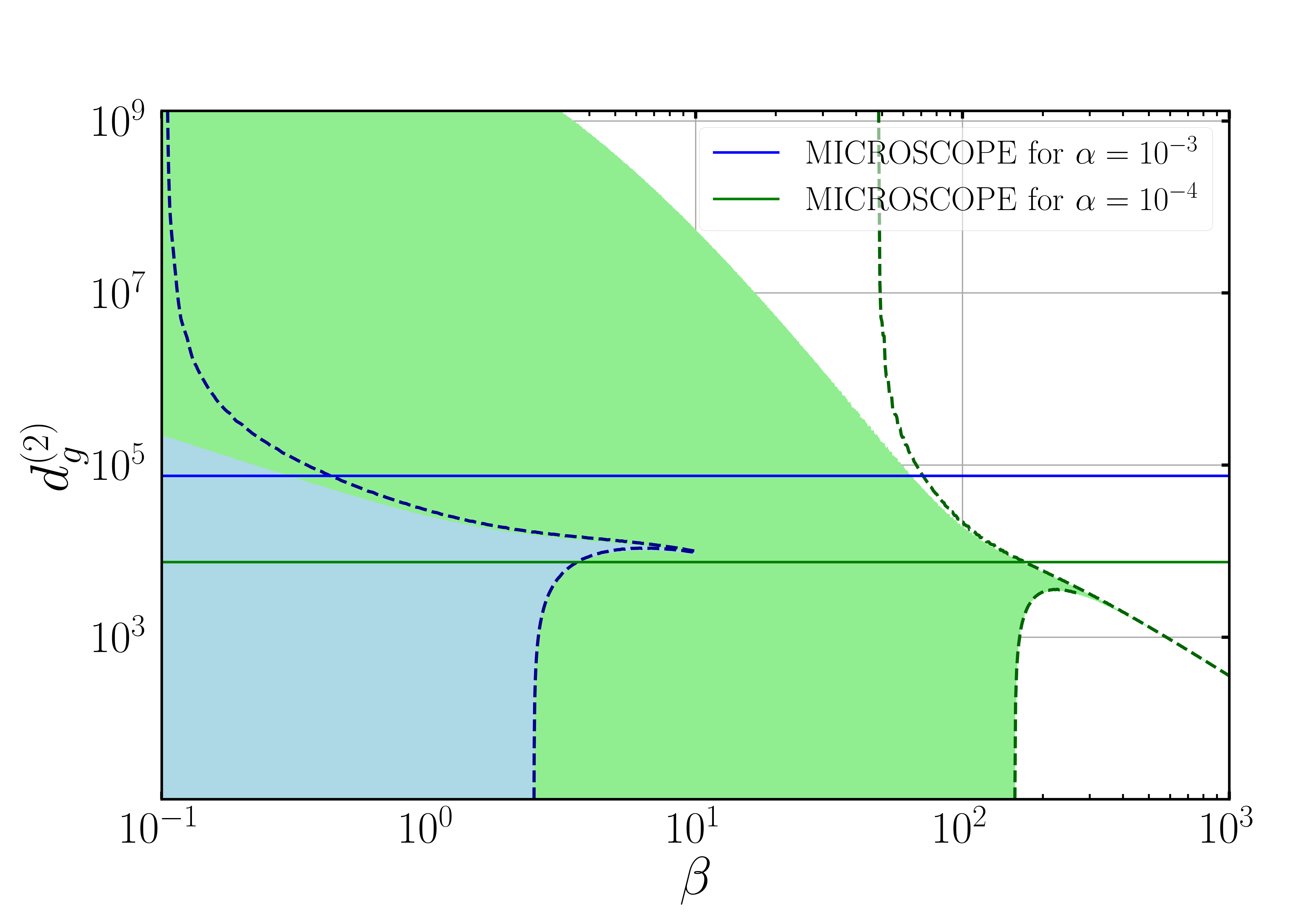}
\caption{Constraints on $1/\Lambda_2$ (top) and $d_g^{(2)}$ (bottom) versus $\beta$ in the cases of $\alpha=10^{-3}$ and $10^{-4}$ in spherical soliton, presented by blue and green regions respectively. {The shaded regions show the parameter space allowed at the 1-$\sigma$ observational level in the point-particle treatment, while the darker blue and green dashed contours mark the corresponding allowed-region boundaries after including the finite-size response of S2.} Constraints from Cassini stochastic gravitational waves measurement \cite{Armstrong:2003ay} and MICROSCOPE experiment \cite{MICROSCOPE:2022doy} are {shown} for comparison. {The regions above the solid comparison lines are excluded.}}
\label{fig:sobeta}
\end{figure}

For the spherical soliton,  similar 2D constraints are shown in Fig.~\ref{fig:sobeta}. The blue and green regions represent the allowed parameter space for \( \alpha = 10^{-3} \) and \( \alpha = 10^{-4} \), respectively. Constraints from Cassini \cite{Armstrong:2003ay} and MICROSCOPE \cite{MICROSCOPE:2022doy} are also included for comparison. The region with \( \beta \gtrsim 1 \) is ruled out for \( \alpha \sim 10^{-3} \), implying that the total mass of the soliton cannot exceed {approximately} the mass of Sgr~A* in this case. {At $\alpha=10^{-3}$ and $\beta\simeq1${, corresponding to a characteristic soliton radius of $\sim0.2\,\mathrm{pc}$,} the dashed boundaries give $1/\Lambda_2\simeq5\times10^{-17}\,\mathrm{GeV}^{-1}$ and $d_g^{(2)}\simeq3\times10^4$. The universal limit is more than one order of magnitude stronger than Cassini and the dilaton limit is stronger than MICROSCOPE.} As \( \alpha \) decreases, the soliton radius increases substantially, rendering the S2 observations less sensitive. {The dashed boundaries agree with the point particle result at weak coupling and give weaker coupling limits at larger coupling.}

\begin{figure}[!ht]
\centering
\includegraphics[width=0.50\textwidth]{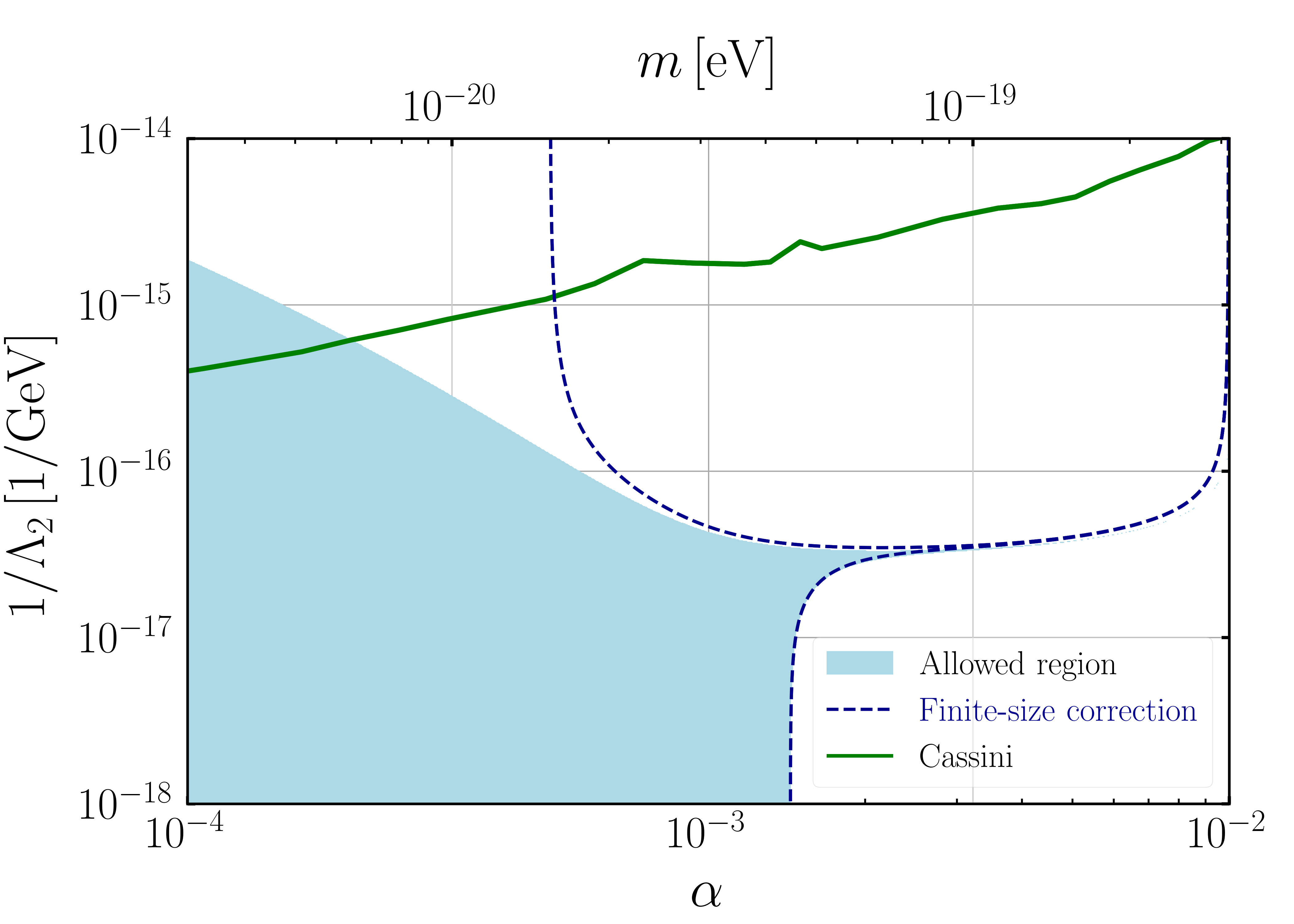}\par
\vspace{1mm}
\includegraphics[width=0.50\textwidth]{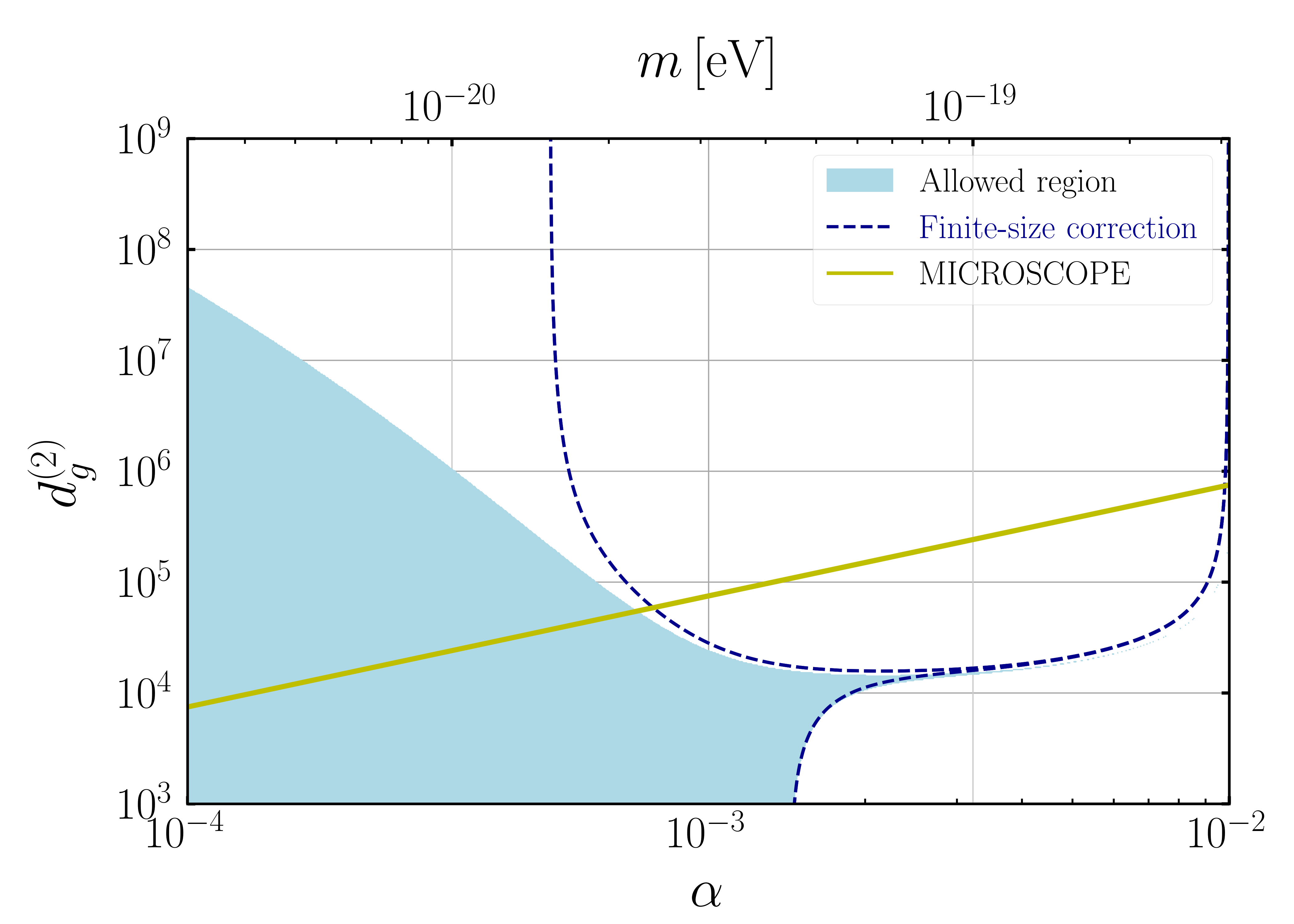}
\caption{{Constraints on $1/\Lambda_2$ (top) and $d_g^{(2)}$ (bottom) versus $\alpha$ for the spherical soliton. The blue shading denotes the allowed region in the point-particle treatment, while the darker blue dashed contours give the corresponding boundaries after the finite-size correction of S2. We adopt the extrapolated halo--soliton relation in Refs.~\cite{Schive:2014dra,Schive:2014hza}. Constraints from Cassini stochastic gravitational-wave measurements~\cite{Armstrong:2003ay} and the MICROSCOPE experiment~\cite{MICROSCOPE:2022doy} are shown for comparison. The regions above the corresponding comparison curves are excluded.}}
\label{fig:so}
\end{figure}

In Fig.~\ref{fig:so}, we show the constraints on \(1/\Lambda_2\) and \(d_g^{(2)}\) as functions of \( \alpha \), together with a comprehensive comparison to results from the Cassini \cite{Armstrong:2003ay} and MICROSCOPE experiments~\cite{MICROSCOPE:2022doy}. {We use the extrapolated halo--soliton relation~\cite{Schive:2014dra,Schive:2014hza},
\begin{equation*}
\beta\simeq\frac{1.06\times10^{-3}}{\alpha}
\left(\frac{M_\mathrm{h}}{1.4\times10^{12}\,M_\odot}\right)^{1/3},
\end{equation*}
where $M_\mathrm{h}$ is the host-halo mass and $M=4.3\times10^6\,M_\odot$ is used for Sgr~A$^*$, as implemented in {Fig.~\ref{fig:so}}. The halo--soliton relation makes $\beta$ increase as $\alpha$ decreases. The low coupling solution requires {$\alpha\lesssim10^{-3}$}. For {$\alpha\gtrsim10^{-3}$}, the gravitational contribution alone is incompatible with the measured precession, and any surviving allowed band is a cancellation window rather than a one sided upper limit. This is consistent with Ref.~\cite{GRAVITY:2024tth}, since the requirement that the total orbit enclosed mass remain below $\sim1000\,M_\odot$ largely excludes the region with $\alpha\gtrsim10^{-3}$. The saturation of $C\mathcal{A}_\star$ sets a common threshold at $\alpha\simeq5.0\times10^{-4}$ in both panels, below which no one sided upper coupling limit remains. After this response is included, the S2 limit is stronger than Cassini for $\alpha\gtrsim5.1\times10^{-4}$, or $m\gtrsim1.6\times10^{-20}\,\mathrm{eV}$, and stronger than MICROSCOPE for $\alpha\gtrsim7.9\times10^{-4}$, or $m\gtrsim2.4\times10^{-20}\,\mathrm{eV}$. The strongest corrected limits occur near {$\alpha\simeq10^{-3}$} and reach $1/\Lambda_2\simeq3.7\times10^{-17}\,\mathrm{GeV}^{-1}$ and $d_g^{(2)}\simeq1.8\times10^4$.}

\FloatBarrier
\section{Summary and outlook}\label{sec:sum}

In this work, we propose using S-stars around Sgr\,A$^*$ to detect scalar ULDM in the GC through both gravitational and non-gravitational interactions. We investigate linear and quadratic couplings between the real scalar field $\phi$ and the baryonic matter of the star, in particular their effects on the stellar orbital dynamics. As an illustration, we consider two possible ULDM structures in the GC, the scalar GA and the spherical soliton. We find a distinct non-oscillatory effect on the orbit in the case of quadratic coupling, which manifests in long-term, secular orbital evolution. Using observational data on the periastron precession of S2, we derive improved constraints on the quadratic coupling constant and the total mass of the scalar cloud in the GC.
In particular, for the GA in $|211\rangle$, we find $\beta\lesssim10^{-3}$ at $\alpha\sim0.04$. At $\alpha\sim0.1$, the first allowed interval extends to $\beta\sim10^{-2}$, with another allowed interval at larger $\beta$ arising from cancellation between the gravitational and non-gravitational contributions to the precession. For the spherical soliton, we find $\beta\lesssim1$ at $\alpha\sim10^{-3}$. {The interpretation of the coupling constraints is governed by the transition between two regimes. At weak coupling, the scalar field penetrates S2 and the point particle result is recovered. At stronger coupling, the finite size response of S2 becomes important. This response becomes appreciable at $1/\Lambda_2\sim4\times10^{-17}\,\mathrm{GeV}^{-1}$ or $d_g^{(2)}\sim2\times10^4$, and the approach of $C\mathcal A_\star$ to a constant makes the non gravitational precession saturate.

The resulting impact depends on the spatial distribution of the scalar field. For the $|211\rangle$ GA at $\alpha=0.04$, the corrected universal and dilaton limits remain stronger than Cassini and MICROSCOPE for $\beta\gtrsim1.5\times10^{-5}$ and $\beta\gtrsim2\times10^{-5}$, respectively. For the spherical soliton at $\alpha=10^{-3}$ and $\beta\simeq1$, the corrected limits reach $1/\Lambda_2\simeq5\times10^{-17}\,\mathrm{GeV}^{-1}$ and $d_g^{(2)}\simeq3\times10^4$. Along the halo--soliton relation, the corrected S2 constraints are stronger than Cassini for $m\gtrsim1.6\times10^{-20}\,\mathrm{eV}$ and stronger than MICROSCOPE for $m\gtrsim2.4\times10^{-20}\,\mathrm{eV}$ within the displayed range. {For $m\gtrsim4.3\times10^{-20}\,\mathrm{eV}$, the parameter space is largely excluded by the gravitational effect of the soliton.} The strongest corrected limits occur near {$\alpha\simeq10^{-3}$}, where $1/\Lambda_2\simeq3.7\times10^{-17}\,\mathrm{GeV}^{-1}$ and $d_g^{(2)}\simeq1.8\times10^4$. Thus S2 provides complementary sensitivity to both universal and composition dependent quadratic interactions when the response of the stellar interior is included.}

Besides the non-oscillatory effect, the oscillatory accelerations induced by both linear and quadratic couplings also affect the star's motion, which can be precisely measured by intensive observations near periapsis. Hence, more specialized, time-dependent analysis is needed for searching oscillatory ULDM signals. From a theoretical perspective, the spectral lines of S-stars can be shifted due to the scalar-Higgs-type couplings~\cite{Yuan:2022nmu} and scalar-photon couplings~\cite{Yuan:2022nmu,bai2025probingaxionsspectroscopicmeasurements}, which are set to zero in this work. Furthermore, while we have assumed a pure two-body system in this work, other perturbations, such as extended mass~\cite{Hees:2017aal,GRAVITY:2024tth} within the orbit of S2 and the external gravitational field, may in reality be affecting. However, their properties remain largely uncertain. A more comprehensive analysis based on observational data of S-stars is therefore warranted. We leave these studies to future investigation.

\appendix

{
\section{Impact of quartic self-interaction}\label{app:self-interaction}
The main analysis neglects scalar self-interactions. A widely studied extension introduces a quartic self-interaction~\cite{Chavanis:2019bnu,Chakrabarti_2022,Dave:2023wjq,banik2025bosonstarshostingblack,Baryakhtar:2020gao,Gavilan-Martin:2026zzw,Witte:2024drg,Helfer_2017,Levkov:2016rkk,arakawa2023detectionbosenovaequantumsensors}. To assess the validity of this approximation for spherical solitons and gravitational atoms (GAs), we consider
\begin{equation}\label{eq:phi:potential}
	V(\phi)=\frac{m^2}{2}\phi^2+\frac{\lambda}{4}\phi^4,
\end{equation}
with $\lambda>0$ ($\lambda<0$) corresponding to repulsive (attractive) self-interaction.

In the nonrelativistic limit, writing
\begin{equation}
	\phi=\frac{1}{\sqrt{2m}}\left(\psi e^{-imt}+\text{c.c.}\right)
\end{equation}
gives the Schr\"odinger--Poisson (SP) (or the Gross-Pitaevskii-Poisson) equations
\begin{align}
	i\partial_t\psi&=-\frac{\nabla^2\psi}{2m}+m\Phi\psi+\frac{3\lambda}{4m^2}|\psi|^2\psi,\\
	\nabla^2\Phi&=4\pi\left[m|\psi|^2+M\delta^3(\mathbf{x})\right],
\end{align}
where $M$ is the black-hole mass and $M_{\rm c}=\int d^3x\,m|\psi|^2$ is the cloud mass. For a spherical stationary state, $\psi=f(r)e^{-iEt}$, the rescaled variables $y=mr$, $V=2(\Phi_{\rm c}-E/m)$, $F=\sqrt{8\pi/m}\,f$, $\alpha=mM$, and $\gamma=3\lambda/(16\pi m^2)$ bring the SP equations to
\begin{align}
	yF''+2F'&=(yV-2\alpha)F+\gamma yF^3,\nonumber\\
	yV''+2V'&=yF^2.\label{app:SP}
\end{align}
For a solution normalized by $F_\kappa(0)=\kappa^2$, the boundary conditions are $F_\kappa'(0)=-\alpha\kappa^2$, $V_\kappa'(0)=0$, and $F_\kappa(\infty)=0$. The scaling
\begin{align}
	F_\kappa(y,\alpha,\gamma)&=\kappa^2F_1(\kappa y,\alpha/\kappa,\kappa^2\gamma),\\
	V_\kappa(y,\alpha,\gamma)&=\kappa^2V_1(\kappa y,\alpha/\kappa,\kappa^2\gamma)
\end{align}
leaves $\Xi\equiv\alpha^2\gamma=\frac{3\lambda M^2}{16\pi}$ invariant. At fixed $(\alpha,\beta,\Xi)$, with $\beta=M_{\rm c}/M$, $\alpha_*$ is determined from
\begin{align}
	\mathcal B(\alpha_*,\gamma_*)&\equiv\frac{1}{2\alpha_*}\int_0^\infty dy\,y^2F_1^2(y,\alpha_*,\gamma_*),\\
	\beta&=\mathcal B\!\left(\alpha_*,\frac{\Xi}{\alpha_*^2}\right),
\end{align}
and $\kappa_*=\alpha/\alpha_*$. This provides the corresponding extension of the parameterization used in Ref.~\cite{yu2025detectingultralightdarkmatter} to nonzero quartic self-interaction.

\begin{figure}[t]
	\centering
	\begin{minipage}[t]{0.485\textwidth}
		\centering
		\includegraphics[width=\linewidth]{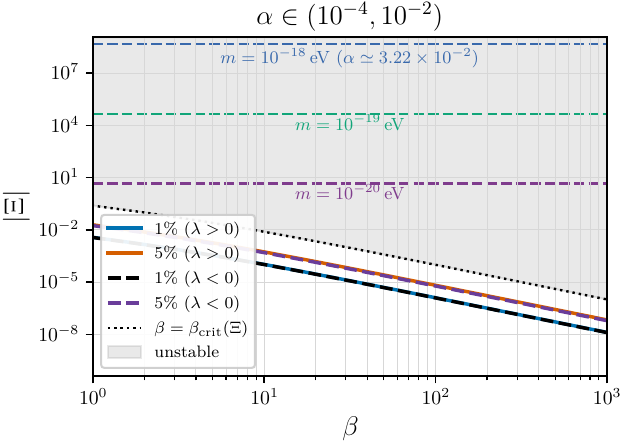}
	\end{minipage}\hfill
	\begin{minipage}[t]{0.485\textwidth}
		\centering
		\includegraphics[width=\linewidth]{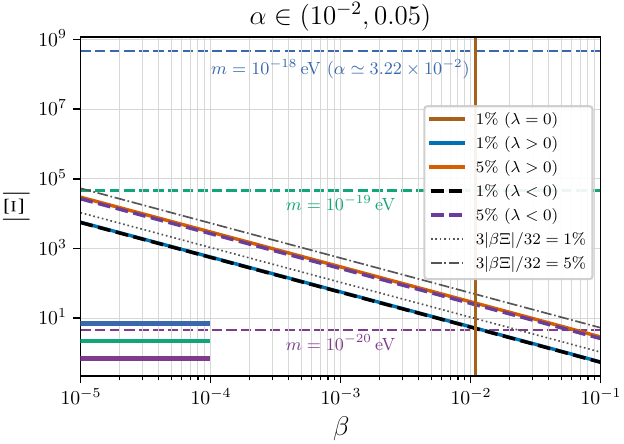}
	\end{minipage}
	\caption{{Effects of quartic self-interaction on the cloud profiles. The left panel shows the $\epsilon_a=1\%$ and $5\%$ contours for spherical solitons with repulsive (solid) and attractive (dashed) interactions for $\alpha\in(10^{-4},10^{-2})$; the dotted curve marks the stability boundary of the attractive branch. The right panel shows the corresponding contours for the self-gravitating solution branch connected to the $|211\rangle$ state for $\alpha\in(10^{-2},0.05)$, together with the perturbative reference lines $3|\beta\Xi|/32=1\%$ and $5\%$ and the contour at which the self-gravitating $\lambda=0$ profile differs from the hydrogenic profile by $1\%$. For each labeled scalar mass, the horizontal dashed line gives the value of $|\Xi|$ implied by the structure-formation constraint in Eq.~\eqref{bound}; in the right panel, the horizontal solid line of the same color marks $|\Xi|_{\rm dyn}$.}}
	\label{fig:quartic-validity}
\end{figure}

Existing bounds set useful reference scales. The Bullet Cluster gives $|\lambda|\lesssim1.7\times10^{-39}(m/10^{-18}\,\mathrm{eV})^{3/2}$~\cite{arakawa2023detectionbosenovaequantumsensors}, while structure formation yields~\cite{Cembranos:2018ulm,arakawa2023detectionbosenovaequantumsensors}
\begin{equation}
	|\lambda|\lesssim5\times10^{-80}\left(\frac{m}{10^{-18}\,\mathrm{eV}}\right)^4.\label{bound}
\end{equation}
For $M=4.30\times10^6\,M_\odot$, Eq.~\eqref{bound} implies $|\Xi|=4.60$, $4.60\times10^4$, and $4.60\times10^8$ at $m=10^{-20}$, $10^{-19}$, and $10^{-18}\,\mathrm{eV}$, respectively. These bounds assume an $\mathcal O(1)$ dark-matter fraction and weaken for a subdominant scalar component.

For the QCD axion\cite{Peccei:1977hh,Weinberg:1977ma,Wilczek:1977pj,PhysRevLett.43.103,DINE1981199,GrillidiCortona:2015jxo}, the zero-temperature NLO chiral perturbation theory result of Ref.~\cite{GrillidiCortona:2015jxo} gives
\begin{align}
	m&=5.70\,\mu\mathrm{eV}\left(\frac{10^{12}\,\mathrm{GeV}}{f_a}\right),\\
	\lambda_{\rm QCD}&=-0.0577\frac{m^2}{f_a^2}\\
	&=-1.78\times10^{-105}\left(\frac{m}{10^{-18}\,\mathrm{eV}}\right)^4,
\end{align}
The associated quadratic nucleon coupling is~\cite{Okawa:2021fto}
\begin{equation}\label{eq:axion}
	\frac{1}{\Lambda_2}\simeq2.2\times10^{-26}\,\mathrm{GeV}^{-1}\left(\frac{m}{10^{-18}\,\mathrm{eV}}\right).
\end{equation}
The QCD-axion quartic coupling corresponds to
\begin{equation*}
\Xi_{\rm QCD}=-1.525\times10^{-19}
\left(\frac{\alpha}{0.01}\right)^4
\left(\frac{M}{4.30\times10^6\,M_\odot}\right)^{-2},
\end{equation*}
which remains far below the scale of appreciable profile deformation throughout
the GA parameter space considered in the main text.

\paragraph{Spherical soliton.}
To assess the validity of the $\lambda=0$ approximation, we characterize the self-interaction-induced change in the gravitational acceleration along the S2 orbit by
\begin{align}
	\mathcal K[a]&\equiv\int_0^{2\pi}df\,r(f)^2|\cos f|\,|a(f)|,
	\\
	\epsilon_a(\Xi,\beta)&\equiv\max_{\alpha\in A}
	\frac{\mathcal K[a_{\text{g},\Xi}-a_{\text{g},0}]}{\mathcal K[a_{g,0}]},
\end{align}
where $r(f)$ is the unperturbed S2 orbit and $A$ denotes the range of $\alpha$ shown in Fig.~\ref{fig:quartic-validity}. The quantity $\epsilon_a$ provides a positive-definite measure of the deformation of the acceleration profile. For attractive self-interactions, the stationary solutions become unstable above a critical cloud mass; the corresponding boundary is shown in Fig.~\ref{fig:quartic-validity}.

\paragraph{Gravitational atom.}
For the self-gravitating branch continuously connected to the $|211\rangle$ state, we evaluate the same diagnostic for an equatorial S2 orbit. The stationary profiles are obtained by solving the SP equations numerically at fixed $(\beta,\Xi)$. At $\Xi=0$, the numerical eigenvalue agrees well with analytic weak-self-gravity result at small $\beta$,
\begin{equation}
	\epsilon\equiv\frac{E}{m\alpha^2}
	=-\frac18-\frac{237}{1280}\beta+\mathcal{O}(\beta^2).
\end{equation}
For weak quartic self-interaction, first-order perturbation theory about the hydrogenic $|211\rangle$ state gives the binding-energy shift
\begin{equation}
	\delta\epsilon_\lambda=\frac{3}{256}\beta\Xi,
	\qquad
	\left|\frac{\delta\epsilon_\lambda}{\epsilon_0}\right|
	=\frac{3|\beta\Xi|}{32},
	\qquad
	\epsilon_0=-\frac18.
\end{equation}
The corresponding $1\%$ and $5\%$ levels are shown in the right panel of Fig.~\ref{fig:quartic-validity} as references. For the nonlinear calculation, we restrict the scan to $\alpha\in(10^{-2},0.05)$. The same panel also shows $|\Xi|_{\rm dyn}=3.91\sqrt{\alpha/0.01}$~\cite{Baryakhtar:2020gao}, the scale at which self-interaction may begin to affect the superradiant dynamics.

Using the structure-formation bound in Eq.~\eqref{bound}, the profile deformation reaches $\epsilon_a=1\%$ at $\beta\simeq1.2\times10^{-2}$ for $m=10^{-20}\,\mathrm{eV}$. The QCD-axion self-interaction is many orders of magnitude weaker, so the $\lambda=0$ approximation remains well justified in that case.
}


{
\section{Finite-size response of S2}\label{app:finite-size}

We estimate the finite-size effect with a uniform-density stellar model,
\begin{align}
\rho_\star=\frac{3M_\star}{4\pi R_\star^3}.
\end{align}
Across S2, the incident scalar field varies negligibly because $R_\star$ is much smaller than both the spatial variation scale of the scalar cloud and the scalar Compton wavelength. {Writing the local real scalar field as $\phi(t,r)=\operatorname{Re}[e^{-i\omega_\phi t}\phi_\star(r)]$, denoting its incident amplitude by $\phi_0$, and using $|m^2-\omega_\phi^2|R_\star^2\ll1$, the leading spatial amplitude $\phi_\star(r)$} in the presence of a positive quadratic matter coupling $C$ obeys
\begin{align}
\nabla^2{\phi_\star}=\begin{cases}
\mu_\star^2{\phi_\star}, & r<R_\star,\\
0, & r>R_\star,
\end{cases}
\qquad
\mu_\star^2=C\rho_\star,
\qquad
z\equiv\mu_\star R_\star.
\end{align}
Here $C=1/\Lambda_2^2$ for the universal coupling, whereas $C=4\pi d_g^{*(2)}\simeq4\pi\times0.907d_g^{(2)}$ when only $d_g^{(2)}$ is nonzero in the dilaton model. Requiring regularity at the stellar center, {$\phi_\star\rightarrow\phi_0$} far from the star, and continuity of {$\phi_\star$} and {$d\phi_\star/dr$} at the surface gives
\begin{align}
{\phi_{\star,\mathrm{in}}}(r)&={\phi_0}\frac{\sinh(\mu_\star r)}{\mu_\star r\cosh z},\\
{\phi_{\star,\mathrm{out}}}(r)&={\phi_0}\left[1-\frac{R_\star}{r}\left(1-\frac{\tanh z}{z}\right)\right].
\end{align}

The effective scalar charge of the extended star is
\begin{align}
Q_\star
&=C\int_{r<R_\star}d^3x\,\rho_\star{\phi_{\star,\mathrm{in}}}(r)
=4\pi R_\star{\phi_0}\left(1-\frac{\tanh z}{z}\right),
\end{align}
while the point-particle value is $Q_\mathrm{pp}=CM_\star{\phi_0}$. Their ratio defines the finite-size response factor used in Figs.~\ref*{fig:ga}--\ref*{fig:so},
\begin{align}
\mathcal{A}_\star(z)
\equiv\frac{Q_\star}{Q_\mathrm{pp}}
=\frac{3}{z^2}\left(1-\frac{\tanh z}{z}\right)
=\frac{3(z-\tanh z)}{z^3},
\end{align}
so that $\overline{\delta\mathbf{a}}_\mathrm{n}^\mathrm{FS}=\mathcal{A}_\star\overline{\delta\mathbf{a}}_\mathrm{n}$. At weak coupling,
\begin{align}
\mathcal{A}_\star(z)=1-\frac{2}{5}z^2+\mathcal{O}(z^4),
\end{align}
and the point-particle limit is recovered. At strong positive coupling,
\begin{align}
\mathcal{A}_\star(z)&=\frac{3}{z^2}\left[1-\frac{1}{z}+\mathcal{O}(e^{-2z})\right],\\
\lim_{C\rightarrow\infty}C\mathcal{A}_\star
&=\frac{3}{\rho_\star R_\star^2}
=\frac{4\pi R_\star}{M_\star}.
\end{align}
Thus the scalar field is increasingly excluded from the stellar interior as the positive coupling grows: $\mathcal{A}_\star$ decreases, but the product multiplying the point-particle acceleration approaches a constant. This produces the high-coupling saturation displayed by the finite-size-corrected contours. The uniform-density treatment is intended as a simple estimate; a realistic stellar density profile would refine the transition without changing this qualitative saturation behavior.
}


\acknowledgments
We thank Jinyi Shangguan and Yu Cheng for useful discussions, and
Gregory Herczeg for constructive comments on the manuscript. This work was
supported by the National Natural Science Foundation of China (12573042), the
Beijing Natural Science Foundation (1242018), the National SKA Program of China
(2020SKA0120300), the Max Planck Partner Group Program funded by the Max Planck
Society, and the High-Performance Computing Platform of Peking University.

\paragraph{Data and code availability.}
The numerical data and code supporting the findings of this study are available
from the corresponding author upon reasonable request.

\bibliographystyle{JHEP}
\bibliography{main}
\end{document}